# On the Raman Spectrum of Plutonium dioxide: vibrational and crystal field modes

M. Naji[1], N. Magnani[1], L. J. Bonales[3], S. Mastromarino[1, 2], J.-Y. Colle[1], J. Cobos[3], D. Manara[1,*]

[1]*European Commission, Joint Research Centre, Institute for Transuranium Elements, P.O. Box 2340, 76125 Karlsruhe, Germany.*
[2]*Università degli Studi di Roma "La Sapienza", Piazzale Aldo Moro 5, 00185 Rome, Italy.*
[3]*Centro de Investigaciones Energéticas, Medioambientales y Tecnológicas, CIEMAT Avenida Complutense, 40, 28040 Madrid, Spain.*

The Raman spectrum of plutonium dioxide is studied both experimentally and theoretically. Particular attention has been devoted to the identification of high-energy modes at 2120 cm$^{-1}$ and 2625 cm$^{-1}$, whose attribution has so far been controversial. The temperature dependence of both modes suggests an electronic origin. Crystal Field (CF) calculations reported in this work shows that these two modes can be respectively assigned to the $\Gamma_1 \rightarrow \Gamma_5$ and $\Gamma_1 \rightarrow \Gamma_3$ CF transitions within the $^5I_4$ manifold. These two modes, together with the only vibrational line foreseen by the group theory for the Fm-3m PuO$_2$ symmetry – the T$_{2g}$ Pu-O stretching mode observed at 478 cm$^{-1}$ – can thus be used as Raman fingerprint of fcc plutonium dioxide.

## I. INTRODUCTION

Plutonium dioxide spectroscopy can have important applications in nuclear waste management, nuclear forensic and safeguard activities. [1, 2] In particular, the Raman spectrum of plutonium dioxide (PuO$_2$) has been studied in a few research laboratories in the last couple of decades. [3-6] Plutonium dioxide crystallises in the fcc fluorite-like structure, belonging to the Fm-3m space group. The factor group theory at the Γ point for this type of crystal symmetry was fully derived by Shimanouchi et al. [7]. It foresees only one triply degenerate T$_{2g}$ (LO1) Raman active vibrational mode, and one infrared active T$_{1u}$ mode at a lower energy. Since the work by Shimanouchi et al., these modes have been experimentally observed for various actinide dioxides and other fluorite-like materials,[8-11] including plutonium dioxide. [4, 5, 12] However, the Raman spectrum of plutonium dioxide was reported up to higher Raman shift energies (beyond 1500 cm$^{-1}$) only recently.[12] A few crucial features of the high-energy part of the PuO$_2$ Raman spectrum remain to-date unclarified, which has not permitted a satisfactory definition of a precise Raman fingerprint for PuO$_2$. Among those features, the most remarkable are two strong peaks consistently observed by Sarsfield et al.[12] at 2120 ± 10 cm$^{-1}$ and 2620 ± 10 cm$^{-1}$. Despite some attempts, the nature of these two lines could not be clearly defined. Sarsfield et al. [12] analysed in depth various possible attributions for them, both in terms of higher-order phonon combinations and surface impurity modes, but could not find any satisfactory assignment. Finally, these authors suggested that these lines might have an electronic origin.

In this respect, there is a vast literature discussing the crystal field (CF) potential of UO$_2$ and NpO$_2$, particularly in relation to their peculiar low-temperature magnetic properties.[13] Less attention has been devoted to PuO$_2$, mainly because, at a first sight, its temperature-independent magnetic susceptibility[14] is compatible with its $\Gamma_1$ singlet ground state (which is expected from the simple argument that the CF states for Pu$^{4+}$ ions in a given ligand geometry should roughly be the inverse as those for U$^{4+}$).[15] However, this single-ion picture turns out to be problematic upon further inspection, in particular because the energy gap between the two lowest CF levels detected by inelastic neutron scattering (INS) is much smaller than the value which was estimated from magnetic measurements (in order to maintain a constant paramagnetic susceptibility up to 1000 K); this issue is still an open problem today. It must be remarked that INS is only sensitive to the $\Gamma_1 \rightarrow \Gamma_4$ transition, and a complete spectroscopic investigation of the lowest $^5I_4$ manifold is still lacking and would be highly desirable in order to test it against current theories.

Experimentally, in addition to INS, electronic Raman scattering (ERS) played a crucial role to determine the various split LSJ manifolds of the *4,5f* ions.[16] In UO$_2$, ERS elucidate the manifestation of the strong electron–phonon interactions in the resonant coupling between nLO (n = 1–4) phonons and $3H'_4 \rightarrow 3F'_2$ U$^{4+}$ intermultiplet CF excitations. Consequently, one would expect that the Raman effect would be of some interest in the study of CF transitions in PuO$_2$.

The experimental part of this study was carried out with the help of an original encapsulation technique, described in a very recent publication.[17] Among other things, this technique permits, for the investigation of such highly radioactive materials, the use of different excitation laser sources, and the employment of the triple subtractive spectrometer configuration. The first feature is important for the analysis of resonant and energy dispersive modes and for the detection of fluorescence features in the Raman spectra; the second is necessary for the study of low-energy Raman lines and anti-Stokes peaks. In particular, the study of the T$_{2g}$ anti-Stokes band, carried out in this research for the very first time on PuO$_2$ in parallel to the Stokes spectrum, allowed us to reasonably estimate a temperature dependence of the observed Raman modes, especially the above-mentioned high-energy bands. Moreover, CF calculations performed in the present research soundly support the attribution of these high-energy lines to PuO$_2$ CF transitions. Such a conclusion, together with the already sound theoretical and experimental description of the vibrational part of the Raman spectrum, allows us to propose a consistent Raman fingerprint for plutonium dioxide.

## II. EXPERIMENTAL METHODS

### A. Samples

The current Raman spectroscopy measurements were performed on a solid polycrystalline PuO$_2$ disk, prepared following the procedure described by De Bruycker et al.[18] The starting material was based on beads obtained by gel-supported precipitation (SOL–GEL). A disk-shaped sample of 8–9 mm in diameter and about 1 mm in thickness was obtained using a bi-directional press. The samples were then sintered in an atmosphere of Ar + H$_2$ with 1500 ppm of H$_2$O to obtain dense material. In order to obtain stoichiometric material (O/M=2.00) and remove the defects produced during the fabrication, the PuO$_2$ sample was subjected to two consecutive heat treatments under a flow of air at 1423 K for 8 hours. Already after the first heat treatment the measured weight gain of the sample corresponded to stoichiometric PuO$_2$ separately measured by thermo-gravimetry (TG). Since, in addition, no further change was noticed after annealing the samples for a second time, it was considered that PuO$_{2.00}$ stoichiometry had been reached. A lattice parameter of 0.5396(1) Å was measured by X-ray diffraction (XRD), corresponding to the literature value for PuO$_2$.[19] The average crystallite size measured by XRD and estimated with the help of scanning electron microscopy (SEM) on a polished surface, was (5 ± 3) μm.

The isotopic composition of the plutonium employed was checked by High Resolution Gamma Spectroscopy (HRGS) to be 93.4 wt.% $^{239}$Pu and 6.4 wt.% $^{240}$Pu. Traces (<0.2 wt.%) of $^{241}$Pu and $^{241}$Am stemming from β-decay of $^{241}$Pu were detected by Thermal Ionisation Mass Spectrometry (TIMS).

### B. Raman measurements

A smaller fragment (about 5 x 1 x 1) mm$^3$ of the PuO$_2$ sample was encapsulated in α-shielding Plexiglas container equipped with a quartz window through which micro-Raman spectra were recorded. The whole procedure has been described in details elsewhere.[17] The Raman microscope was equipped for the present measurements with a long working distance (10.6 mm) objective which offers a 0.5 numerical aperture with x50 magnification. The Raman spectrometer employed in this research is a Jobin-Yvon T 64000 equipped with a 1800 grooves per mm grating, a low noise liquid nitrogen cooled symphony CCD detector, a subtractive pre-monochromator (in triple mode) which allows access to anti-Stokes spectrum lines while blocking the elastic Rayleigh line. Excitation sources used in this work are the 488 nm (2.53 eV) or 514.5 nm (2.41 eV) lines of an Ar$^+$ Coherent® Continuous Wave (CW) laser or the 647 nm (1.91 eV) or 752 nm (1.65 eV) lines of a similar Kr$^+$ laser. Both laser sources have a controllable nominal power up to 1.5 W. Typical laser power values at the sample surface ranged between 5 mW and 300 mW, whereby the highest power values were used in order to heat the sample *in situ*. The power impinging the sample surface was measured by a Coherent® power-meter placed at a position corresponding to the sample surface. It is lower by approximately a factor 5 than the actual power at the exit of the laser cavity. Using the long focal x50 objective and the single spectrometer mode permits a good spectral resolution (± 1 cm$^{-1}$) independently of the surface shape, with a spatial resolution of 2 μm x 2μm on the sample surface. The spectrograph is calibrated with the T$_{2g}$ excitation of a silicon single crystal, set at 520.5 cm$^{-1}$.[20] The instrument is calibrated on a daily basis prior to measurements.

### C. Temperature determination

#### a) Stokes and anti-Stokes Raman peak analysis

For Raman laser heating experiments, the Raman laser was used as an excitation and heating source. The laser power was then increased stepwise and the Stokes/anti-Stokes Raman spectra were collected at different powers.

The absolute temperature (T$_{SAS}$) was derived from the ratio (R$_{SAS}$) between the experimental intensities of the anti-Stokes and Stokes T$_{2g}$ peaks according to Bose-Einstein statistics. Prior to these calculations, Stokes and anti-Stokes Raman spectra were corrected for the instrumental response according to the procedure described elsewhere.[21,22] Despite the instrumental correction, the value of T$_{SAS}$ can represent the real sample temperature in a rather inaccurate fashion, due to uncontrollable sample-dependent factors affecting the intensities of the Stokes and anti-Stokes lines, such as impurity fluorescence, roughness scattering etc. Therefore, the value of T$_{SAS}$ should be considered as only indicative of the absolute temperature produced by the laser beam on the irradiated spot. Nonetheless, the values obtained through Stokes and anti-Stokes are sufficient to perform a qualitative study of the temperature evolution of the Raman lines observed in this work.

#### b) The thermodynamic Grüneisen parameter

Another method to estimate the temperature is based on the definition of the Grüneisein parameter γ.[23] It describes the relative shift of phonon frequencies as a consequence of changing the volume of the crystal lattice. In $PuO_2$, at the $\Gamma$ (q ~ 0) point the dispersion curve of $\gamma$ for the $T_{2g}$ mode is flat enough to be considered as a constant $\gamma$ ~1.5.[24, 25]

In the current approximation, a direct correlation between $\gamma$ and the temperature increase induced by a higher power of the laser beam is provided by the thermal expansion of the investigated material. Thus, a linear relation between the $\Delta\tilde{v}_{T2g}(T)$ band Raman shift and the temperature variation can be easily deduced from the definition of the Grüneisen parameter for the $T_{2g}$ mode $\Delta\tilde{v}_{T2g}(T) = \tilde{v}_{T2g0} \cdot 3 \cdot \alpha \cdot \gamma_{T2g} \cdot \Delta T$ , with $\tilde{v}_{T2g0}$ is the phonon wavenumber (478 cm$^{-1}$) measured at 298K and α is the linear thermal expansion coefficient. This latter has been approximated to $1 \cdot 10^{-5}$ K$^{-1}$ according to recent literature values.[26]

At low temperature, both methods (SAS intensity ratio (1) and Grüneisen parameter (2)) converge to reasonable values. With temperature increasing, the temperature difference diverges. This can be due mainly to the over-estimation of temperature in method (1) and to the phonon-phonon interactions that dominates the downshift of the $T_{2g}$ mode in method (2). Thus, at high temperature we believe that the current method (Grüneisen parameter) is not valid to account for the relative shift and mainly anharmonic effects are responsible for the frequency downshift. This has been observed in $UO_2$, where anharmonicity is responsible for the frequency redshift at reasonably low temperatures.[27]

Finally, the two approaches allowed us to give a reasonable estimate of the local sample temperature by averaging the values obtained with the two procedures, based on utterly independent physical effects.

### III. RESULTS AND DISCUSSION

#### A. *Raman spectra of $PuO_2$ with different excitation sources*

Figure 1 shows the Stokes Raman spectra measured on $PuO_2$ with the four excitation sources used in this research. The acquisition of Raman spectra recorded at 752 nm and 647 nm laser excitations had to be cut well below 3000 cm$^{-1}$ because the spectral limit of the CCD detector was reached. Some peaks are clearly observable, which are summarised in Table I.

The $T_{2g}$ peak at (478 ± 2) cm-1 dominates the spectra recorded with all the excitation sources. It corresponds to the LO1 phonon at the centre (or $\Gamma$ point) of the Brillouin zone. A weaker mode is observable in all the spectra at ~578 cm$^{-1}$. This mode has been assigned to the $T_{1u}$ LO2 phonon at the $\Gamma$ point of the Brillouin zone, in agreement with the density functional theory (DFT) calculations performed by Zhang et al.[28] and an original many-body approach proposed by Yin and Savrasov.[24] Such vibrational mode would be Raman forbidden in a perfect Fm-3m crystal. It is activated by Frenkel-type oxygen defects, as suggested in recent literature for an analogous mode predicted and observed in $UO_2$.[29-31] The first overtone (2LO2) of this latter line can be observed at 1156 cm$^{-1}$ in the spectra excited with the higher energy sources, 2.41 eV and 2.53 eV. The assignment of this overtone is based on the exhaustive analysis of a similar peak observed by many authors in $UO_2$.[10] The fact that the 2LO2 line is only visible at higher excitation energies suggests that in $PuO_2$, like in $UO_2$, the mode is activated via a resonance multi-phonon process, its absorption edge lying between 2 eV and 2.4 eV. The deconvolution of this line is difficult in the 488 nm-spectrum because it overlaps with a further broad band centred at 956 cm$^{-1}$. This band might be similar to the weak one observed by Livneh and Sterer in $UO_2$,[10] and assigned to disorder-activated $2TO_R$ phonon at the Brillouin zone's X-edge. However, Yin and Savranov attributed it to the first overtone of the LO1 phonon ($T_{2g}$ line) mode.[24] In fact, this band is centred at exactly twice the energy of the $T_{2g}$ band, whereas its larger width can be attributed to an enhanced disorder effect observable in the overtone. A similar behaviour is observed in the 2LO2 band centred at 1156 cm$^{-1}$. On the other hand, it is interesting to note that its frequency coincides with the $\Gamma_1 \rightarrow \Gamma_4$ (CF) transition, as observed by INS experiment and predicted by DFT and CF calculations.[32-34] The overlapping of this band with the 2LO2 mode especially at high temperatures makes the analysis of its temperature dependence rather difficult and its assignment to a pure $2TO_R$, 2LO1 mode or $\Gamma_1 \rightarrow \Gamma_4$ $Pu^{4+}$ CF excitation ambiguous.

Other broad bands can be seen in the 488-nm and the 514-nm spectra between 1500 cm$^{-1}$ and 2000 cm$^{-1}$. These can be attributed to further overtones (e.g.: 2LO1 + LO2=1531 cm$^{-1}$), adsorbed $O_2$ molecules (1555 cm$^{-1}$), a slight carbon contamination (near 1600 cm$^{-1}$) and fluorescence effects (broad band centred 1800 cm$^{-1}$ in the 488-nm spectrum only). The weak carbon and adsorbed oxygen lines would be visible at higher excitation energies due to the absolute frequency dependence $(\omega - \omega_0)^4$ of the Raman cross section. Moreover, the attribution to overtones of bands centred at spectral positions corresponding to integer multiples of known modes is consistent with the fact that they are only observable with higher energy excitation sources.

Beyond 2000 cm$^{-1}$ only three excitation sources could be used, because the 752 nm laser line is too close to the high-wavelength spectral edge of the spectrometer. In the three cases, two peaks can be obviously observed at 2110 cm$^{-1}$ and 2620 cm$^{-1}$, though less easily in the 647-nm spectrum. These peaks were already observed and studied by Sarsfield et al.[12], although no satisfactory attribution was found for them. Thanks to their extensive analysis, however, several possible assignments were positively discarded, like the combination of phononic overtones or the adsorption of $O_2$, $N_2$, $H_2O$, CO or $CO_2$ on the $PuO_2$ surface. By comparison with the broadly studied Raman spectrum of graphite, we can also exclude that such peaks are linked to surface carbon impurities in $PuO_2$. The typical

Raman peaks of graphite in this spectral range, remeasured in this work under the same conditions employed for PuO$_2$, do lie at 2328 cm$^{-1}$, 2440 cm$^{-1}$ and 2725 cm$^{-1}$, too far away from the current bands. The only possibility left open by Sarsfield et al. was indeed that those peaks have an electronic origin, a hypothesis which we are further testing in the present work.

It can be noticed, incidentally, that a photon energy-dependence of the intensity of these two bands is most likely to be present also in this case. However, such an observation should be treated with care, because no obvious photon-energy dependence was observed by Sarsfield et al. with other experimental facilities and excitation sources for the 2620 cm$^{-1}$ peak, whereas a slight trend could be suggested for the 2110 cm$^{-1}$ one. Therefore, the higher intensities of the two latter peaks observable in the 488-nm spectrum might be linked to the presence of the above-mentioned background fluorescence. Nonetheless, a resonant process can probably be assumed also in this case, certainly for the 2110 cm$^{-1}$ band.

### B. Temperature dependence of PuO$_2$ Raman bands

SAS Raman spectra obtained with the 514-nm excitation for various laser power levels are shown in Figure 2. Prior to the Stokes and anti-Stokes spectra collection, the sample was irradiated for several minutes with a high laser power (about 1 W at the exit of the cavity), in order to locally anneal in situ possible self-radiation damage effects in the investigated spot. The excitation source yielding the best accuracy for the current Stokes and anti-Stokes (SAS) Raman analysis was the 514-nm laser line. Lower energy excitation sources were observed to result, even at the highest power levels, in a very limited sample heating and in any case in a low signal-to-noise ratio in the anti-Stokes spectrum. On the other hand, a non-negligible fluorescence background, clearly observable also in Figure 1, would affect the analysis in spectra recorded around the 488-nm laser line.

Figures 3 a-c display the resulting temperature evolution of the spectral peak position $\Delta\omega/\omega_0 = \frac{\omega(T)-\omega_{298K}}{\omega_{298K}}$, full width at half maximum FWHM $\frac{\Delta\Omega}{\Omega_0} = \frac{\Omega(T)-\Omega_{298K}}{\Omega_{298K}}$ and intensity (normalised as follow, $I(T_{2g}) = I(T_{2g})/(I(T_{2g}) + I(2110) + I(2620))$ for the three main Raman modes observed in the current spectra, the T$_{2g}$ peak and the two higher energy peaks centered at 2110 cm$^{-1}$ and 2620 cm$^{-1}$. These bands were fitted, after baseline correction, with unconstrained Lorentzian curves.

The intensity of the T$_{2g}$ band increases with temperature, up to around 500 K, at which it saturates and stays approximately constant. The full width at half the maximum (FWHM) of this mode moderately increases as a function of T, due to an increased disorder, and its peak position slightly decreases due to thermal expansion, according to the Grüneisen parameter value reported above. The absolute intensity behaviour of this line follows a pure Bose-Einstein statistics only up to the saturation temperature. The Debye temperature $T_D = \frac{hc\tilde{\nu}_\Gamma}{k_B}$ of the LO1 phonon being 686 K, at the current temperatures a monotonic increase of both the Stokes and anti-Stokes intensities is expected. The saturation behaviour observed here is certainly linked to phonon annihilation processes already observed for other oxides, such as, for example, ZnO.[35] Such processes result in the observed peak broadening with increasing temperature (Figure 3 c).

The temperature dependence of the peaks at 2110 cm$^{-1}$ and 2620 cm$^{-1}$ is essentially different, both for the intensity and the widths (Figures 3a and 3c). The intensity of these two modes slightly decreases at the beginning of the investigated temperature range, then stabilises starting from approximately 400 K. Since the widths of these peaks is larger than the one observed for the T$_{2g}$ mode and no clear temperature trend can be deduced (Figure 3c), it is possible that at these temperatures phonon scattering has already largely reduced the lifetime of these excitations. Independently of such effects, a constant intensity of these two bands in the investigated temperature range would be consistent with an electronic origin.

On the other hand, the spectral peak positions of these two modes decreases at increasing temperature, much more than observed for the T$_{2g}$ band. Correspondingly, a much larger value of the Grüneisen parameter can be obtained for the two high-energy peaks, in both cases between 15 and 20. Such a high value of $\gamma$ is probably meaningless, if the bands under investigation are to be considered as purely vibrational modes, unless a very large anharmonicity is assumed for them.

Finally, the markedly different temperature dependence of the high-energy modes, when compared with the purely vibrational T$_{2g}$ line, proves that these two modes are of an indeed electronic nature as already suggested by Sarsfield et al.[12]

In order to complete the assignment of these modes, in this work PuO$_2$ CF energy levels have been recalculated with an updated set of parameters.

It can be readily noticed that the temperature evolution of the T$_{2g}$ band (relative to a purely vibrational mode) is essentially different from the trends observed for the two bands at 2110 cm$^{-1}$ and 2620 cm$^{-1}$.

### IV. CRYSTAL FIELD CALCULATIONS

Krupa and Gajek[34] exploited ab-initio calculations in order to establish the tendency of the evolution of the CF potential as a function of the atomic number for the actinide dioxide series, thus obtaining a set of parameters for PuO$_2$ from experimental spectroscopic studies of the isostructural UO$_2$ and NpO$_2$. In a similar fashion, Kern et al.[32] used Newman's superposition model[36] to determine how the UO$_2$ CF parameters would rescale to PuO$_2$. In both cases, the calculated position of the $\Gamma_4$ triplet is about 930 cm$^{-1}$ above the $\Gamma_1$ ground singlet; this is in good agreement with inelastic neutron scattering (INS) experiments on polycrystalline

$^{242}$PuO$_2$, which have revealed a single wide peak around 1000 cm$^{-1}$ that must be assigned to the intramultiplet $\Gamma_1 \rightarrow \Gamma_4$ transition (the only one allowed by the magnetic-dipole selection rules which govern INS).[33] The parameter set used in the latter paper also predicts correctly the absolute value of the neutron cross section.

This interpretation leaves one problem open: the energy gap value measured by INS is way smaller than the lower bound estimated from the measured magnetic susceptibility.[14] For this reason, Colarieti-Tosti et al.[33] performed a LDA ab-initio study of the electronic structure of PuO$_2$ and included antiferromagnetic exchange enhancement; however, this is not sufficient to completely remove the discrepancy. The CF splittings that they calculate are somewhat smaller with respect to the experimental data, which is also true for other, more recent ab initio studies.[37, 38] All of these papers also predict that the order of the $\Gamma_3$ and $\Gamma_5$ excited levels belonging to the ground $^5I_4$ manifold is reversed with respect to Krupa and Gajek's results;[34] we attribute this to the relatively close position of the $\Gamma_5$ triplet belonging to the excited $^5I_5$ manifold (at an energy around 5000 cm$^{-1}$), which mixes with the lower $^5I_4$-$\Gamma_5$ triplet and pushes it down in energy when the CF is stronger. This consideration emphasizes the need to correctly estimate the CF parameters and, in particular, to fully account for $J$ mixing in the calculations.[39]

The failure to reconcile the existing bulk and spectroscopic measurements for PuO$_2$ is an extremely complicated problem to solve, to the point that recent first-principle calculations even questioned the validity of the single-ion approximation itself in describing the electronic ground state of PuO$_2$.[40] Conversely, in this Section we show that calculations performed with the most reliable values of the CF parameters reproduce the spectroscopically-determined energy splitting of the low-energy manifold very well and are perfectly consistent with an electronic attribution of the two Raman peaks detected at 2110 cm$^{-1}$ and 2620 cm$^{-1}$. The appropriate single-ion Hamiltonian for a tetravalent Pu ion in a CF potential of cubic symmetry is

$$H = \sum_{k=2,4,6} F^k f_k + \zeta_{5f} \sum_{n=1}^{4} \mathbf{s}_n \cdot \mathbf{l}_n + \alpha G(R_3) + \beta G(G_2)$$
$$+ \gamma G(R_7) + \sum_{i=2,3,4,6,7,8} T^i t_i$$
$$+ \sum_{j=0,2,4} M^j m_j + \sum_{k=2,4,6} P^k p_k$$
$$+ B_4 \left[ C_0^{(4)} + \sqrt{\frac{5}{14}} \left( C_{+4}^{(4)} + C_{-4}^{(4)} \right) \right]$$
$$+ B_6 \left[ C_0^{(6)} - \sqrt{\frac{7}{2}} \left( C_{+4}^{(6)} + C_{-4}^{(6)} \right) \right],$$

where $F^2$, $F^4$, and $F^6$ are the Slater radial integrals quantifying the non-spherical part of the electronic repulsion; $\zeta_{5f}$ is the spin-orbit coupling constant for the $5f$ shell; α, β, and γ are the Trees parameters associated to two-body operators which perturbatively account for multiconfigurational effects; the $T^i$ parameters are related to effective operators which describe three-body interactions; the Marvin parameters $M^j$ account for further relativistic corrections such as spin-spin and spin-other-orbit coupling; $P^k$ represent the electrostatically-correlated magnetic interactions; and finally, $B_4$ and $B_6$ quantify the cubic CF potential.[36] For the reasons stated above, we consider that the best estimate for the two latter parameters of PuO$_2$ is that given by Kern et al.[32] as $B_4 = -9760$ cm$^{-1}$ and $B_6 = +3968$ cm$^{-1}$. All other parameters which appear in $H$ are related to free-ion interactions and, therefore, practically independent of the host matrix; for this reason, we have fixed their values to those determined by Carnall et al.[41] for tetravalent Pu ions in PuF$_4$.

The energy levels for the lowest manifold calculated by a full diagonalization of $H$ (in order to correctly take both intermediate coupling and $J$ mixing into account) are reported in Table II, together with the irreducible representation (irrep) of each level (as for the degeneracy of electronic multiplets, we recall that the $\Gamma_1$ state is a singlet, $\Gamma_3$ a doublet, $\Gamma_4$ and $\Gamma_5$ are triplets). Experimental spectroscopic data are now available for the whole manifold ($^5I_4$); focusing on that, it can be noticed that not only the energy experimentally measured by INS for the $\Gamma_1 \rightarrow \Gamma_4$ transition is correctly reproduced by the single-ion model, but also that the predicted $\Gamma_1 \rightarrow \Gamma_5$ and $\Gamma_1 \rightarrow \Gamma_3$ electronic transition energies are in very good quantitative agreement with the position of the Raman peaks measured in this work. Both the latter transitions are allowed by the quadrupolar selection rules, which justify their presence in our unpolarised Raman measurement. This correlation clarifies the origin of the as yet unassigned features in the Raman spectrum and completes the general picture of the CF levels within $^5I_4$. Since over this energy range, only pure CF excitations are expected and no phonon or multi-phonon process is expected, as compared to UO$_2$, both bands should be exclusively attributed to pure CF excitations.

At the end, it is of importance to note that these modes were observed also for PuO$_2$-ThO$_2$ and PuO$_2$-UO$_2$ solid solutions,[6, 42] and their intensity increases with the increase of the plutonium content. This feature is very helpful to identify PuO$_2$ when mixed or dissolved in other compounds, since the vibrational T$_{2g}$ mode, usually overlap with other active vibrations in the phonon spectral region.

## V. CONCLUSIONS

In the present investigation we have described and discussed the main features of the Raman spectrum of plutonium dioxide. The characteristics of the peaks recorded at 2110 cm$^{-1}$ and 2620 cm$^{-1}$ have been broadly discussed in relation to their temperature dependence. Their most plausible attribution is an electronic nature. In particular, their spectral positions match well the $\Gamma_1 \rightarrow \Gamma_5$ and $\Gamma_1 \rightarrow \Gamma_3$ crystal electric field transitions calculated in this work. This correlation clarifies the origin of the as yet unassigned features in the Raman spectrum and completes the general picture of the CF levels within the $^5I_4$ manifold. The reported peaks at 2110 cm$^{-1}$ and 2620 cm$^{-1}$ and their attributions define a peculiar Raman fingerprint of plutonium dioxide, which can be used for the in-situ and ex-situ detection of this important compound for nuclear safety and safeguard. As a closing remark, we reiterate that the single-ion model can naturally describe all the experimental spectroscopic features of $PuO_2$ within the same framework that works well for the whole actinide dioxide series;[39] this will have to be taken into account in future attempt to theoretically describe its electronic ground state.[43]

**Acknowledgment**

M.N would like to acknowledge the European Commission for the financial support. M. N and D. M would like to thank Pr. R. Caciuffo and Pr. R. Konings (JRC, ITU) for reading our manuscript. We also would like to thank Drs. P. Simon (CEMHTI-CNRS) and D. Neuville (IPGP, CNRS) for their fruitful discussions.

# FIGURES

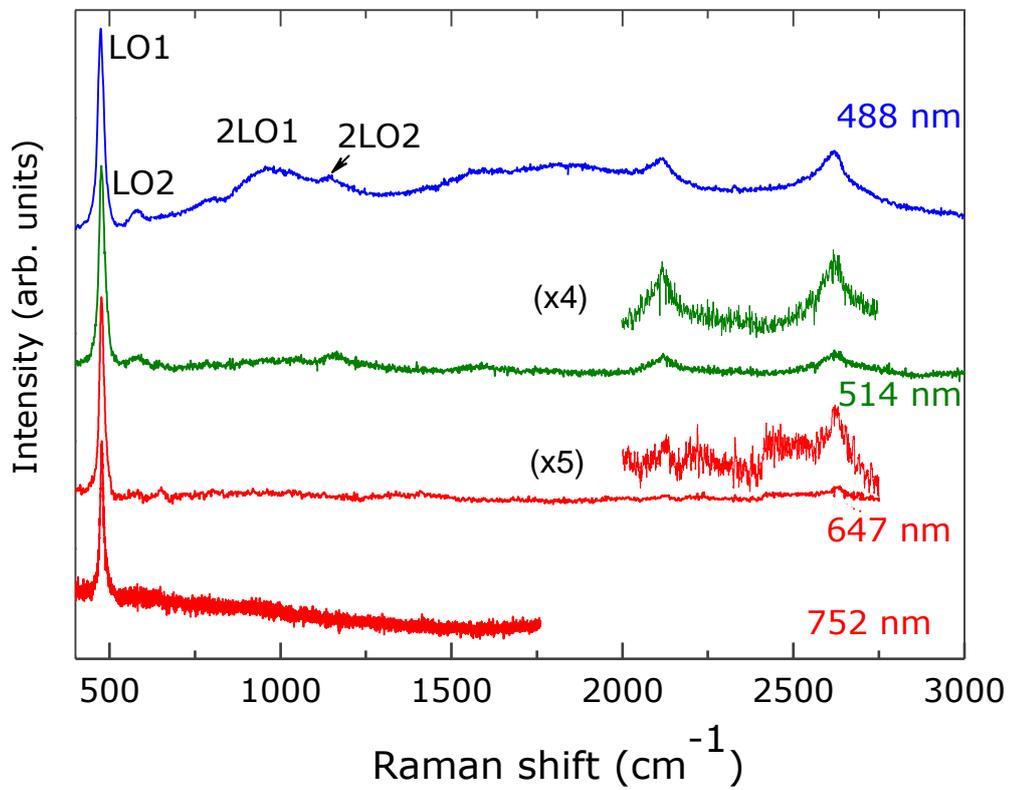

FIG.1. (Color online) Spectra of PuO$_2$ acquired with different laser energies 488 nm (2.53 eV), 514.5 nm (2.41 eV), 647 nm (1.91 eV), 752 nm (1.65 eV). All bands are assigned in accordance with Table I (a) and (b).

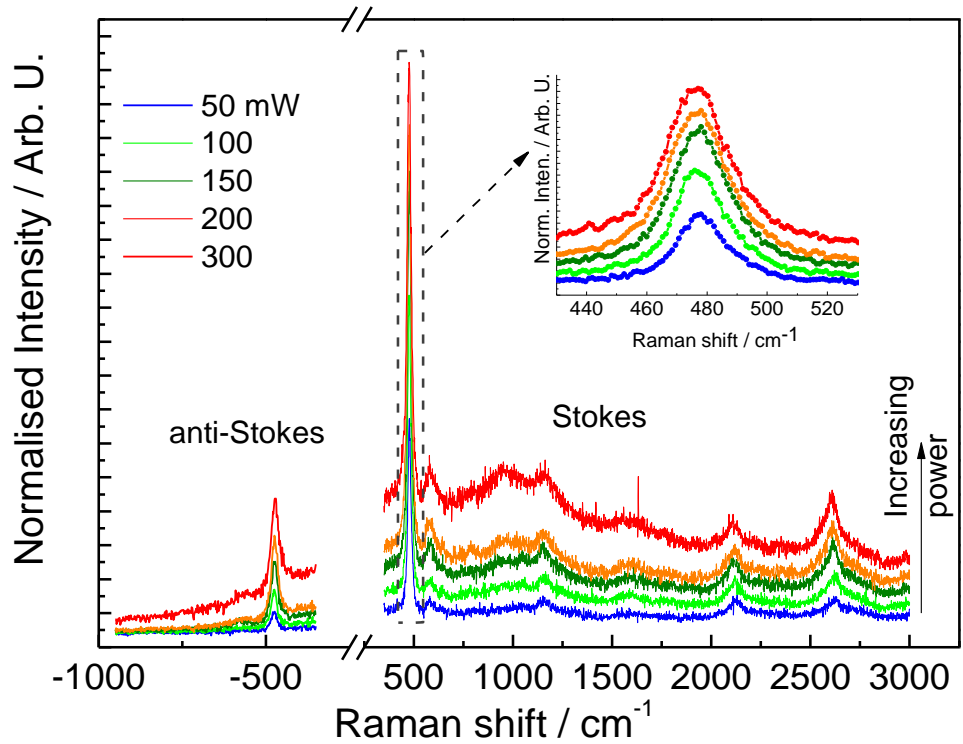

FIG.2. (Color online) Stokes and anti-Stokes Raman spectra obtained with 514 nm (2.41 eV) for different laser power. The solid-line arrow shows the increasing power from 50 to 300 mW. The inset shows the $T_{2g}$ peak shift with increasing the laser power. Laser power is recorded at the sample surface.

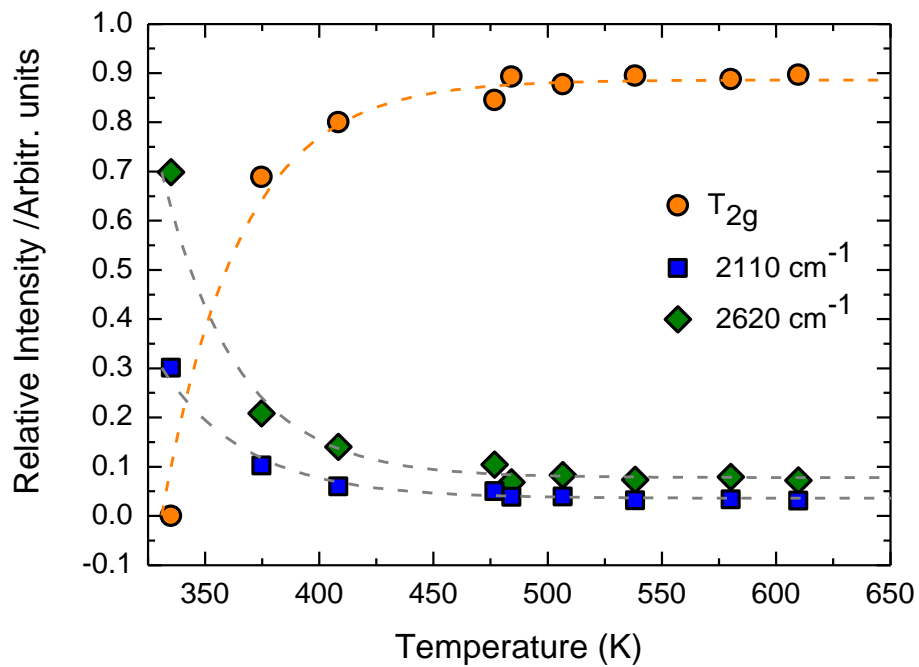

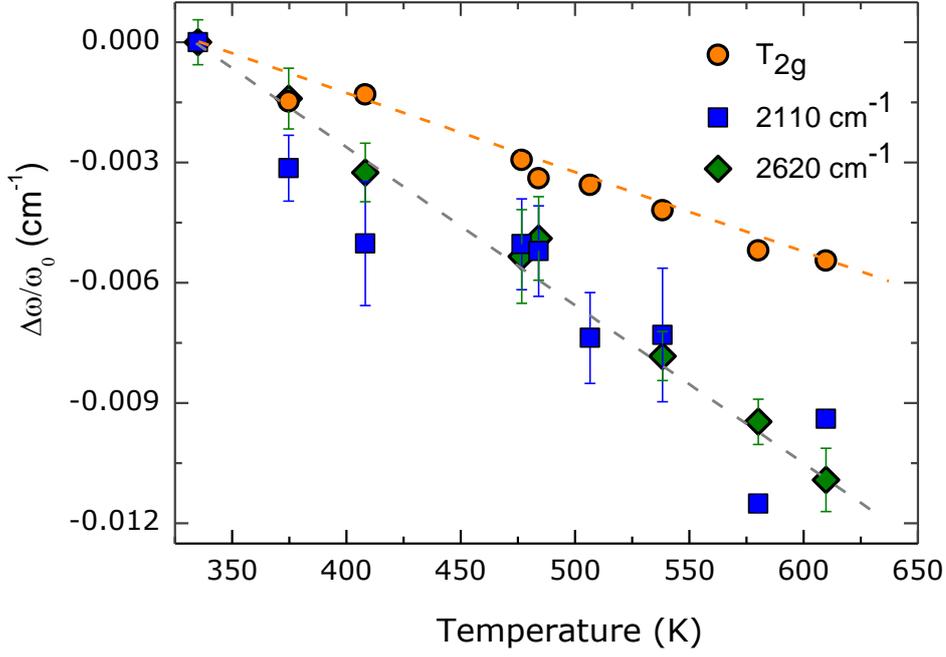

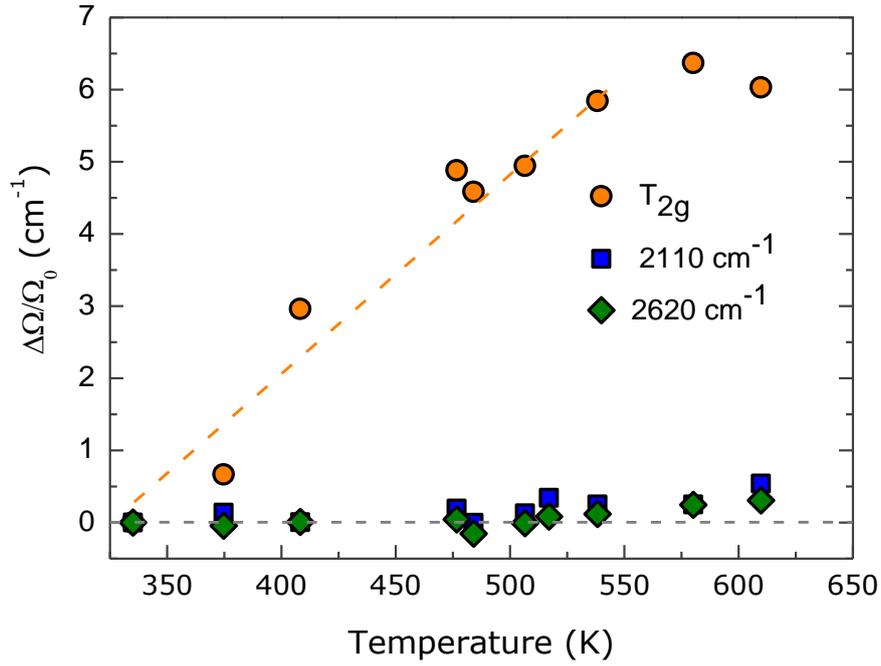

FIG. 3: (Color online) (a) Temperature dependence of intensity $I(T_{2g}) = I(T_{2g})/(I(T_{2g}) + I(2110) + I(2620))$, (b) peak shift $\Delta\omega/\omega_0 = \frac{\omega(T)-\omega_{298K}}{\omega_{298K}}$ and (c) FWHM $\frac{\Delta\Omega}{\Omega_0} = \frac{\Omega(T)-\Omega_{298K}}{\Omega_{298K}}$ of $T_{2g}$ (filled circles), 2110 cm$^{-1}$ (filled squares), 2620 cm$^{-1}$ (filled diamonds) bands. Dashed lines in (a) represent a linear fit of the peak position, and in (b,c) are guides to the eye.

# TABLES

**TABLE I.** Attribution of the Raman band position for $PuO_2$

| Raman band position (cm$^{-1}$) | Assignment | Reference |
|---|---|---|
| 478±2 | LO1($T_{2g}$) | 12 |
| 575 | LO2 | 12 |
| 956 | 2TO$_R$ or 2LO1? | This work |
| 1156 | 2LO2 | 12 |
| 1531 | 2LO1+ LO2 | This work |
| 1555 | O$_2$ | 44 |
| 1600 | sp$^2$(C=C) | 45 |
| 1800[a] | Fluorescence | This work |
| 2110[b] | $\Gamma_1 \rightarrow \Gamma_5$ (CF) | This work |
| 2620[b] | $\Gamma_1 \rightarrow \Gamma_3$ (CF) | This work |

[a] Visible only with the 488 nm laser line
[b] Not observed with the 752 nm laser line.

**TABLE II:** Comparison between the calculated and measured CF energy splitting (expressed in cm$^{-1}$) for the lowest manifold of $PuO_2$, from the present work as well as literature sources. The energy of the lowest level ($\Gamma_1$) has been set to zero in all cases.

| irrep | Calculated energies | | | | | | Measured energies | |
|---|---|---|---|---|---|---|---|---|
| | Ref. [34] | Ref. [32] | Ref. [33] | Ref. [37] | Ref. [38] | This work | Ref. [32] | This work |
| $\Gamma_4$ | 929 | 928 | 799 | 518 | 782 | 1055 | 1000 | – |
| $\Gamma_5$ | 1776 | – | 1678 | 1024 | 1645 | 2153 | – | 2110 |
| $\Gamma_3$ | 2153 | – | 1307 | 829 | 1573 | 2630 | – | 2620 |